\documentclass[12pt]{article}

\usepackage[final]{graphics}
\usepackage[dvips]{graphicx}
\usepackage{epsfig}
\usepackage{amsmath, amsthm, amssymb}
\usepackage{cite}
\numberwithin{equation}{section}

\begin{document}

\title{ {\huge Bulk black holes radiating in non-$Z_2$ brane-world spacetimes}}

\author{
David Jennings$^1$\footnote{D.Jennings@damtp.cam.ac.uk}, 
Ian R. Vernon$^2$\footnote{I.R.Vernon@dur.ac.uk},\\
Anne-Christine Davis$^1$\footnote{A.C.Davis@damtp.cam.ac.uk},
Carsten van de Bruck$^3$\footnote{C.vandeBruck@sheffield.ac.uk} \\
\normalsize \em $^1$Department of Applied Mathematics and Theoretical Physics,\\
\normalsize \em CMS, University of Cambridge, Cambridge, CB3 0WA, UK. \\
\normalsize \em $^2$Centre for Particle Theory, University of Durham, \\
\normalsize \em South Road, Durham, DH1 3LE, U.K. \\
\normalsize \em $^3$Department of Applied Mathematics,\\
\normalsize \em University of Sheffield, Sheffield, S3 7RH, U.K.\\
}
\maketitle

\begin{abstract}
In this paper we present a general asymmetric brane model involving arbitrary energy transport to and from an embedded 4-D FRW universe. We derive a locally defined mass function for the 5D spacetime and describe its time evolution on the brane. We then specialise our model to the two cases of graviton production in the early universe and radiating black holes in the bulk.
\end{abstract}



\newpage

\section{Introduction}
The idea of having extra, unseen, spatial dimensions to our universe has its origins with 
Kaluza and Klein \cite{kaluza,klein} who proposed a five dimensional version of General Relativity 
that incorporated electromagnetism. Their idea was to include a small, compact, extra dimension. 
Our inability to observe this extra dimension was attributed to its scale being outside our 
experimental reach. The notion of small, hidden extra dimensions later reappeared with the 
advent of string theory and superstring theory where their existence was necessary for the 
finiteness of the theories.\\
The modern extension of string theory is M-theory, where, in addition to one dimensional 
strings, we have higher dimensional objects called branes. Brane-worlds spring from a model 
suggested by Horava and Witten \cite{hwI,hwII} where the strong coupling limit of the 
$E_8\times E_8$ heterotic string theory at low energy is described by eleven dimensional 
supergravity. An important new element in this model is that one of the extra dimensions, 
the orbifold dimension, can actually be quite large. The reason for this is that the 
standard model particles are represented by open strings while gravitons are represented 
by closed strings, these open strings are constrained to finish on a brane while the 
closed strings have complete freedom. In brane-world 
scenarios~\cite{ruba,akam1,anton,visser,ahdd,ahddII,aahdd,sund1,rscompact99,rshierarchy99}
our universe is then viewed as a 3-brane embedded in a larger spacetime with one extra 
dimension which is not necessarily compact (see~\cite{carrev,langrev,royrev} for recent reviews). 
The standard model particles are attached 
to this brane, while the gravitons are free to move off it into the extra dimension, 
which is referred to as the \textit{bulk}. 

The usual model consists of a bulk Einstein-Hilbert action together with a brane action,
\begin{eqnarray}
 S_{\mbox{\tiny EH \normalsize}} &=& -\int dx^5 \sqrt{-g_5} \left ( \frac{R}{2 \kappa^2} + \Lambda \right ),\nonumber\\
S_{\mbox{\tiny brane \normalsize}} &=& \int dx^4 \sqrt{-g_4} (-V) ,
\end{eqnarray}
where $V$ is the brane tension for our 3-brane universe and $\Lambda $ is the bulk 
cosmological constant. As such the brane is viewed as a hypersurface in the 5-D spacetime. 
Its induced metric and extrinsic curvature are defined as
\begin{eqnarray}
 h^A_B &=& \delta ^A_B - n^An_B,\nonumber\\
K_{AB} &=& h^C_Ah^D_B \nabla_Cn_D , 
\end{eqnarray}
where $n_A$ is a vector field normal to the hypersurface and capital indices run over the five dimensions.
Since we assume that the standard model particles are confined to the brane, the 
matter contribution from our universe will take the form of a delta function in $T^A_B$. 
Then, using Einstein's equations we may obtain the famous Darmois-Israel junction 
conditions \cite{s22} relating the jump in the extrinsic curvature across a 
hypersurface to its matter distribution on the hypersurface,
\begin{eqnarray}
 [K_{AB}] = -\kappa ^2 \left ( T_{AB,\mbox{\tiny brane \normalsize}} - \frac{1}{3}h_{AB}T_{\mbox{ \tiny brane \normalsize}} \right ) .
\end{eqnarray}
The spatial components of the Israel-junction conditions for the brane produce a 
modified FRW equation for the Hubble parameter~\cite{noncosmo,bulkcosmo,cosmo1extrad,expans1extrad},
\begin{eqnarray}
 H^2 = \left ( \frac{\dot{a}}{a}\right )^2 = \frac{8 \pi \mbox{G}}{3} \rho \left ( 1 + \frac{\rho}{2V} \right ) + \frac{\Lambda_4}{3} - \frac{k}{a^2} + \frac{C}{a^4} ,
\end{eqnarray}
where we have assumed that the brane universe is homogeneous and isotropic and that 
a $Z_2$-symmetry is present across the brane.
This differs from the standard FRW equation in two ways: firstly we now have a 
term \emph{quadratic} in the density, whose presence would dominate in the very early universe; 
secondly there is a term $\frac{C}{a^4}$ which is referred to as the dark radiation and which 
would also become relevant in the early stages of the universe. The dark radiation term is 
actually the one remaining component of the projected Weyl tensor \cite{maeda1} for the bulk 
spacetime and so is an external feature determined by the content of the bulk, e.g. 
gravitons flowing in the extra dimension.
This extra term could provide 
a possible test for the above extra-dimensional scenarios. Although many authors have assumed 
that $C$ is constant, this will not be the case if there is energy flowing to, 
form  or across the brane. 

In this paper we develope a general formalism to describe the dynamics of an isotropic, homogeneous 
non-$Z_2$ (or asymmetric) 3-brane in a bulk that possesses arbitrary energy 
transport to, from and across the brane. We then 
use these results to examine a variety of situations, including the case where the energy transport is due to 
the bulk black holes either side of the brane emitting Hawking 
radiation, in addition to the brane thermally emitting gravitons. 
The paper is organised as follows: in section~\ref{secnz2} we summarise the important aspects of 
non-$Z_2$ symmetric branes, before presenting the necessary formalism to determine their 
evolution in a general bulk. We derive the Friedmann equation along with the necessary equations for 
the time-dependent dark radiation parameter $C$ and non-$Z_2$ parameter $F$, then we discuss 
the locally defined mass function, related in some cases to the black hole mass. 
In section~\ref{secsb} we apply our general equations to the cases of graviton emission by the brane and to 
Hawking radiation by both black holes, before presenting a numerical treatment and discussion of the various 
cases in section~\ref{secna}. Our conclusions are given in section~\ref{secc}.

\section{Non-$Z_2$ branes}\label{secnz2}
In this section we develop the equations governing energy transport to and from a non-$Z_2$ 
brane universe, where we do not identify the two bulks either side of the 
brane~\cite{Kraus,tyerscosmo,branevsshell,fpaper}. 
We then 
specialise to the case of radial null fluids in the bulk. The presence of several overlapping 
null fluids in the bulk makes the overall description problematic and an analysis using 
Vaidya metrics \cite{lang,Leeper,Leeper2} will not work. The natural coordinate system to 
use would seem to be a double null system of ingoing and outgoing null rays, but this 
presents several technical difficulties which makes analysis awkward. It is convenient for 
our purposes to began with the Einstein equations in 5-D, using a coordinate system based 
on the brane - as opposed to tackling issues of extrinsic curvature in a bulk-based system. 
 \subsection{General description }
We work with a five dimensional metric describing a standard 3-brane with the usual 
homogeniety and isotropy for its spatial dimensions,
\begin{eqnarray}\label{metric}
 ds^2 = -n^2(t,z)dt^2 + a^2(t,z)\gamma_{ij}dx^idx^j + b^2(t,z)dz^2 ,
\end{eqnarray}
and a total 5D energy-momentum tensor $T^A_B$ which we decompose as,
\begin{eqnarray}
 T^A_B =T^A_{B, \mbox{\tiny brane, vac \normalsize}} + T^A_{B, \mbox{\tiny bulk, vac \normalsize}} +  T^A_{B, \mbox{\tiny brane, matter \normalsize}} + T^A_{B,\mbox{\tiny bulk, matter \normalsize}} .
\end{eqnarray}
We shall assume separate cosmological constants on the brane and in the bulk, and a perfect fluid on the brane. At present we shall make no assumptions concerning the energy-momentum tensor in the bulk. 
\begin{eqnarray}
 T^A_{B, \mbox{\tiny brane, vac \normalsize}} &=& \frac{\delta(z)}{b} \mbox{diag}(-V,-V,-V,-V,0),\nonumber\\
 T^A_{B, \mbox{\tiny bulk, vac \normalsize}\pm} &=&\mbox{diag}(-\frac{\Lambda_\pm}{\kappa^2},-\frac{\Lambda_\pm}{\kappa^2},-\frac{\Lambda_\pm}{\kappa^2},-\frac{\Lambda_\pm}{\kappa^2},-\frac{\Lambda_\pm}{\kappa^2}), \nonumber\\
 T^A_{B, \mbox{\tiny brane, matter \normalsize}} &=& \frac{\delta(z)}{b} \mbox{diag}(-\rho,p,p,p,0) .
\end{eqnarray}
Here the $\pm$ refer to the two bulk spacetimes either side of the brane and we shall work without the usual $Z_2$-symmetry across the brane. To aid the discussion we define, for any quantity $X$, its jump and average across the brane,
\begin{eqnarray}
 \Delta X = X_{0^+} - X_{0^-} ,\nonumber\\
 \overline{X} = (X_{0^+} + X_{0^-})/2 ,
\end{eqnarray}
The Einstein equations
\begin{eqnarray}
 G_{AB} = \kappa^2 T_{AB} ,
\end{eqnarray}
have $G_{AB}$ given by~\cite{noncosmo,bulkcosmo},
\begin{eqnarray}
 G_{00} &=& 3\left ( \frac{\dot{a}}{a}\left ( \frac{\dot{a}}{a} + \frac{\dot{b}}{b} \right ) - \frac{n^2}{b^2} \left (\frac{a''}{a} + \frac{a'}{a} \left (\frac{a'}{a} - \frac{b'}{b} \right ) \right ) + k \frac{n^2}{a^2} \right ) ,\nonumber\\
 G_{ij} &=& \frac{a^2}{b^2}\gamma_{ij} \left ( \frac{a'}{a}\left ( \frac{a'}{a} + 2\frac{n'}{n} \right ) - \frac{b'}{b} \left (\frac{n'}{n} + 2\frac{a'}{a}\right ) + 2\frac{a''}{a} + \frac{n''}{n} \right )+ {} \nonumber\\
& & {} + \frac{a^2}{n^2}\gamma_{ij}\left (-\frac{\dot{a}^2}{a^2} + 2\frac{\dot{a}\dot{n}}{an} - 2\frac{\ddot{a}}{a} -2 \frac{\dot{a}\dot{b}}{ab} +\frac{\dot{n}\dot{b}}{nb} -\frac{\ddot{b}}{b} \right ) -k\gamma_{ij} ,\nonumber\\
G_{05} &=& 3 \left ( \frac{n'\dot{a}}{na} + \frac{a'\dot{b}}{ab} -\frac{\dot{a}'}{a} \right ) ,\nonumber\\
 G_{55} &=& 3\left ( \frac{a'}{a}\left ( \frac{a'}{a} + \frac{n'}{n} \right ) - \frac{b^2}{n^2} \left (\frac{\ddot{a}}{a} + \frac{\dot{a}}{a} \left (\frac{\dot{a}}{a} - \frac{\dot{n}}{n} \right ) \right ) - k \frac{b^2}{a^2} \right ) .
\end{eqnarray}
\subsection{Active source equation - The jump conditions}
We may use the fact that only the second derivatives with respect to $z$ contain delta functions and integrate the $(00)$-equation and the $(ij)$-equation over the small interval $[ -\epsilon, +\epsilon ]$, which contains the brane at $z=0$. This provides us with the jump conditions across the brane in terms of the matter content on the brane~\cite{noncosmo},
\begin{eqnarray}
 \Delta n' &=& \frac{\kappa^2}{3}n_0b_0(3p +2\rho -V) ,\nonumber\\
 \Delta a' &=& -\frac{\kappa^2}{3}a_0b_0(\rho+V) ,
\end{eqnarray}
where the primes denote derivatives with respect to the transverse coordinate $z$, and $X_0$ is the quantity $X$ evaluated on the brane at $z=0$. For convenience we go to the temporal gaussian coordinates where,
\begin{eqnarray}
 b_0 = n_0 = 1 ,\nonumber\\
\dot{n}_0 = 0 .
\end{eqnarray}
For the sake of clarity, we now drop the subscript $0$. In addition, all unspecified terms $T_{AB}$ shall henceforth refer to the \emph{bulk content}. 
\subsection{Passive source equation}
The Darmois-Israel junction conditions on their own are incomplete when the $Z_2$-symmetry across the brane is removed. It is thus necessary to supplement them with a passive gravitational source equation describing the motion of the brane within the background spacetime. The origin of this passive source is found in the divergence equation for the brane universe \cite{bran2},
\begin{eqnarray}
\nabla_A T^{AB}_{\mbox{\tiny brane \normalsize}}&=& f^B ,
\end{eqnarray}
where $f^B$ is a force density localised on the brane. We then consider the orthogonal component of this field,
\begin{eqnarray}
 f&=&n_Bf^B .
\end{eqnarray}
After some tensor manipulation, the dynamical equation is integrated across the brane and we obtain the passive gravitational source equation,
\begin{eqnarray}
 \overline{T}^{AB}_{\mbox{\tiny brane \normalsize}}\overline{K}_{AB}&=&\overline{f} .
\end{eqnarray}
It has been observed \cite{bran2} that this is analogous to Newton's second law of motion, with $\overline{T}_{AB}$ taking the place of mass and $\overline{K}_{AB}$ being the acceleration.

In our current setting we may obtain the explicit form of the passive equation by analysing the $(55)$-equation. We find that it takes the form,
\begin{eqnarray}
 \overline{n'}(\rho + V) - 3 \frac{\overline{a'}}{a}(p-V) = \frac{\Delta \Lambda}{\kappa^2} -\Delta T_{55} ,
\end{eqnarray}
which for a $Z_2$-symmetric brane is satisfied trivially.
\subsection{Transverse equations}
The $(05)$-equation provides us with the equation governing energy flow off or onto the brane. With the help of the jump conditions, which relate transverse derivatives to the energy content of the brane, we find that this takes the form,
\begin{eqnarray}
 \dot{\rho} + 3H(p + \rho) &=& \Delta T_{05} ,\nonumber\\
H &=& \frac{\dot{a}}{a} .
\end{eqnarray}
This form is quite natural, with the usual energy conservation replaced with one containing a `leaking' term, $\Delta T_{05}$, which corresponds to the net energy flux into the extra dimension.

The average of the $(05)$-equation tells us how the geometrical asymmetry across the brane evolves with energy transfer from one side to the other. We find that,
\begin{eqnarray}\label{average_a}
 \frac{1}{a}\partial_t \overline{a'} = \overline{n'}H - \frac{\kappa^2}{3}\overline{T}_{05},
\end{eqnarray}
which will be needed later.
\subsection{The FRW equation for the brane}
From the $(55)$-equation we may now arrive at an acceleration equation for the brane scale factor,
\begin{eqnarray}\label{acceleration}
 H^2 + \frac{\ddot{a}}{a} + \frac{k}{a^2} &=& \lambda_1 - \left ( \frac{\kappa^2}{6} \right )^2 [ V(3p - \rho) + \rho (3p + \rho) ] + \Theta  - \frac{\kappa^2}{3} \overline{T}_{55} ,\nonumber\\
\lambda_1 &=& \frac{\kappa^2}{3} \left ( \frac{\overline{\Lambda}}{\kappa^2} + \frac{\kappa^2 V^2}{6} \right ) ,\nonumber\\
\Theta &=& \left ( \frac{\overline{a'}}{a} \right )^2 + \frac{\overline{a'}\overline{n'}}{an} .
\end{eqnarray}
Where $\Theta $ is the geometrical asymmetry contribution.
We now \emph{define} the auxiliary field $\chi$, which represents a generalised dark radiation term, via the FRW equation,
\begin{eqnarray}
 H^2 &=& \beta \rho^2 + 2\gamma \rho -\frac{k}{a^2} + \chi + \lambda ,
\end{eqnarray}
where $\beta$,$\gamma$, $\lambda$ are constants. Using this together with equation (\ref{acceleration}) we find that the auxiliary field dynamically obeys,
\begin{eqnarray}
 \dot{\chi} + 4H\chi +2H ( \frac{\kappa^2}{3}\overline{T}_{55} - \Theta ) = -\frac{\kappa^4}{18}(\rho + V )\Delta T_{05} .
\end{eqnarray}
From its form we see, in the absence of asymmetry, that it behaves like radiation coupled to matter on the brane.
Using the acceleration equation for the brane scale factor and the defining equation for $\chi $ we require that the constants in the definition of $\chi$ have the following values,
\begin{eqnarray}
\beta &=& \kappa^4/36 ,\nonumber\\
\gamma &=& \beta V ,\nonumber\\
\lambda &=& \frac{\kappa^2}{6}(\frac{\overline{\Lambda}}{\kappa^2} + \frac{\kappa^2V^2}{6}) .
\end{eqnarray}
It is now possible to split this generalised dark radiation term, $\chi$, into two natural parts. To describe the asymmetric geometry across the brane we define the function $F(t)$ as,
\begin{eqnarray}
 \overline{a'} = \frac{F(t) + \frac{1}{4\kappa^2} \Delta \Lambda a^4}{(\rho + V)a^3} .
\end{eqnarray}
We now decompose $\chi$ into two parts and in the process introduce the Weyl term $C(t)$ which is the usual dark radiation factor in the FRW equation,
\begin{eqnarray}
\chi = \frac{C(t)}{a^4} + \frac{(F +\frac{1}{4\kappa^2} \Delta \Lambda a^4)^2}{(\rho + V)^2a^8} .
\end{eqnarray}
 This gives the FRW equation for a non $Z_2$-symmetric brane in a completely general bulk,
\begin{eqnarray}
 H^2 &=& \beta \rho^2 + 2\gamma \rho -\frac{k}{a^2} + \frac{C(t)}{a^4} + \frac{(F +\frac{1}{4\kappa^2} \Delta \Lambda a^4)^2}{(\rho + V)^2a^8} + \lambda ,
\end{eqnarray}
\subsection{Evolution of parameters}
$F(t)$ represents the asymmetry of energy across our brane boundary, and consequently we expect its development to be decided by transverse flows to the brane. With this in mind we return to equation (\ref{average_a}) and find that,
\begin{eqnarray}
 \dot{F} = \frac{\Delta T_{05}}{\rho + V}(F +\frac{1}{4\kappa^2} \Delta \Lambda a^4) - Ha^4 \Delta T_{55} - \frac{\kappa^2}{3}a^4(\rho + V) \overline{T}_{05} .
\end{eqnarray}
It is then a simple matter to deduce that the Weyl term $C(t)$ must obey,
\begin{eqnarray}
 \! \! \! \!\! \!\frac{3}{2 \kappa^2}\dot{C} = \frac{\overline{T}_{05}}{\rho + V}(F +\frac{1}{4\kappa^2} \Delta \Lambda a^4) - Ha^4 \overline{T}_{55} - \frac{\kappa^2}{12} a^4(\rho + V) \Delta{T_{05}} .
\end{eqnarray}
These equations, together with,
\begin{eqnarray}
 H^2 &=& \beta \rho^2 + 2\gamma \rho -\frac{k}{a^2} + \frac{C(t)}{a^4} + \frac{(F +\frac{1}{4\kappa^2} \Delta \Lambda a^4)^2}{(\rho + V)^2a^8} + \lambda ,\nonumber\\
 \dot{\rho} &=& - 3H(p + \rho) + \Delta T_{05} ,
\end{eqnarray}
constitute a complete set of equations describing the cosmological evolution of the brane-world model. \\
We now make the observation that,
\begin{eqnarray}
 \frac{3}{2 \kappa^2} \dot{C} + \frac{1}{2} \dot{F} = \left (\frac{(F +\frac{1}{4\kappa^2} \Delta \Lambda a^4)}{\rho + V} - \frac{\kappa^2}{6}a^4(\rho + V) \right ) T_{05,+} - Ha^4 T_{55,+} ,\nonumber\\
 \frac{3}{2 \kappa^2} \dot{C} - \frac{1}{2} \dot{F} = \left (\frac{(F +\frac{1}{4\kappa^2} \Delta \Lambda a^4)}{\rho + V} + \frac{\kappa^2}{6}a^4(\rho + V) \right ) T_{05,-} - Ha^4 T_{55,-} ,\nonumber
\end{eqnarray}
and so the two quantities $\frac{3}{2 \kappa^2} C + \frac{1}{2} F$ and $\frac{3}{2 \kappa^2} C - \frac{1}{2} F$ are intrinsic in some way to the $+$ and $-$ bulks respectively. We shall demonstrate that these two quantities correspond to the energy content of their respective bulks, first by considering Vaidya spacetimes and then in a more general setting. 
\subsection{Vaidya spacetimes}
The Vaidya spacetimes \cite{ivaid,gvaid} generalise AdS-Schwarzschild and describe the inflow or outflow of radial null fluids. The metric for such a spacetime takes the form,
\begin{eqnarray}
 ds^2 &=& -f(v,r)dv^2 + 2 \epsilon dv dr + r^2 \gamma_{ij}dx^idx^j ,\nonumber\\
f&=& -\frac{\Lambda}{6}r^2 - \frac{2m(v)}{r^2} ,
\end{eqnarray}
where $v=$ constant describe radial null rays which are ingoing for $\epsilon =1$ and outgoing for $\epsilon =-1$. We may derive the FRW equation for a brane in this spacetime by using the 5-D junction conditions \cite{s22},
\begin{eqnarray}
 [ K_{AB} ] = \kappa ^2 ( T_{AB} - \frac{1}{3} T h_{AB}) ,
\end{eqnarray}
where $h_{AB}$ is the induced metric on the brane. If we do not impose $Z_2$-symmetry we obtain a FRW 
equation~\cite{Leeper2},
\begin{eqnarray}
 H^2 &=& \beta \rho^2 + 2\gamma \rho -\frac{k}{a^2} + \frac{2\overline{m}}{a^4} + \frac{(\frac{3}{\kappa ^2}\Delta m +\frac{1}{4\kappa^2} \Delta \Lambda a^4)^2}{(\rho + V)^2a^8} + \lambda\nonumber,\\
\end{eqnarray}
and so we may infer that,
\begin{eqnarray}
 C&=& 2\overline{m}\nonumber,\\
 \frac{\kappa^2}{3}F &=& \Delta m . 
\end{eqnarray}
We then write the equations governing the evolution of the cosmological parameters as,
\begin{eqnarray}\label{dimensionfull}
H^2 &=& \frac{\kappa ^4}{36}\rho (\rho + 2V) -\frac{k}{a^2} + \frac{2\overline{m}(t)}{a^4} + \frac{(12 \Delta m  + \Delta \Lambda a^4)^2}{16 \kappa ^4(\rho + V)^2a^8} ,\nonumber\\
 \dot{m}_\pm &=& -\left [\pm\frac{\kappa^4}{18}a^4 (\rho + V)  - \frac{\Delta m + \frac{\Delta \Lambda}{12} a^4}{\rho + V} \right ]T_{05,\pm} - \frac{\kappa^2}{3} Ha^4T_{55,\pm} ,\nonumber\\
 \dot{\Delta m} &=& \frac{\Delta T_{05}}{\rho + V}(\Delta m +\frac{\Delta \Lambda}{12} a^4) - \frac{\kappa^2}{3}Ha^4 \Delta T_{55} - \frac{\kappa^4}{9}a^4(\rho + V) \overline{T}_{05} ,\nonumber\\
\dot{\overline{m}} &=& \frac{\overline{T}_{05}}{\rho + V}(\Delta m +\frac{\Delta \Lambda}{12} a^4) - \frac{\kappa^2}{3}Ha^4 \overline{T}_{55} - \frac{\kappa^4}{36}a^4(\rho + V) \Delta{T_{05}} ,\nonumber
\end{eqnarray}
where we have imposed the Randall-Sundrum tuning conditions\cite{rshierarchy99,rscompact99},
\begin{eqnarray}
 \overline{\Lambda } + \frac{\kappa ^4V^2}{6} = 0 ,
\end{eqnarray}
which means the cosmological constant on the brane vanishes.

\subsection{Locally defined mass function}
We now demonstrate that this identification of the bulk mass parameter with linear combinations of $C$ and $F$ may be extended to the case of a general bulk with the usual homogeneous and isotropic brane boundary.
If we begin with the metric (\ref{metric}) and introduce the following Kodama vector field~\cite{kodama},
\begin{eqnarray}
 k^A &=&(\frac{-a'}{nb}, 0, 0, 0, \frac{\dot{a}}{nb}),
\end{eqnarray}
we find that this vector field is conserved:
\begin{eqnarray}
 \sqrt{-g} k^A_{;A}&=&( \sqrt{-g}k^A)_{,A}=0.
\end{eqnarray}
We now use the fact that $T^A_{B;A} = 0$ to define the following conserved current in the bulk spacetime,
\begin{eqnarray}
 J^A&=& \sqrt{-g}T^A_BK^B .
\end{eqnarray}
It is easily checked that $\partial_A J^A =0$ and so we may define a conserved charge $M(t,z)= \int J^0 dz$ satisfying,
\begin{eqnarray}
\dot{M}(t,z)&=& -J^5(t,z) ,
\end{eqnarray}
where $J^5(t,z)$ is the flux of $M$ out through the $z = $constant hypersurface at the point $(t,z)$. Explicitly, in terms of our metric components, this becomes,
\begin{eqnarray}
\dot{M}(t,z) &=& - (T^5_5 \frac{\dot{a}(t,z)}{a} - T^5_0 \frac{a'(t,z)}{a})a(t,z)^4 .
\end{eqnarray}
This expression is now evaluated on the brane from either side. For example,
\begin{eqnarray}
\dot{M}(t,z=0_+) &=& - (T^5_{5,+} H(t) - T^5_{0+} \frac{a_+'}{a})a^4 .
\end{eqnarray}
We then make use of the jump conditons and the definition of the asymmetry function $F$,
\begin{eqnarray}
(a_+' - a_-')/2 &=&-\frac{\kappa ^2}{6}(\rho +V) ,\nonumber\\
(a_+' + a_-')/2 &=& \frac{F(t) + \frac{1}{4\kappa^2} \Delta \Lambda a^4}{(\rho + V)a^3} ,
\end{eqnarray}
to arrive at the result,
\begin{eqnarray*}
 \dot{M}(t,z=0_+) = \left (\frac{(F +\frac{1}{4\kappa^2} \Delta \Lambda a^4)}{\rho + V} - \frac{\kappa^2}{6}a^4(\rho + V) \right ) T_{05,+} - Ha^4 T_{55,+} .
\end{eqnarray*}
Similarly, approaching the brane from the other side we obtain,
\begin{eqnarray*}
 \dot{M}(t,z=0_-) = \left (\frac{(F +\frac{1}{4\kappa^2} \Delta \Lambda a^4)}{\rho + V} + \frac{\kappa^2}{6}a^4(\rho + V) \right ) T_{05,-} - Ha^4 T_{55,-} .
\end{eqnarray*}
We then conclude that the mass function defined locally at each point in the 5D spacetime obeys,
\begin{eqnarray} 
M(t,z=0_\pm)&=& \frac{3}{2 \kappa ^2} C \pm \frac{1}{2} F ,
\end{eqnarray}
which provides us with the physical relation the two components of $\chi$ have with the bulk spacetimes.

\section{Specific bulks}\label{secsb}
We may now specialise our dynamical equations to different bulk contents. In order to describe the system completely we need to specify four independent physical quantities, $T_{05,\pm}$, $T_{55,\pm}$, to describe the effects the two different bulks have on the brane. These are then combined together into $\Delta T_{05}$, $ \Delta T_{55}$, $ \overline{T}_{05}$, $ \overline{T}_{55}$.

If we consider a null fluid with energy momentum tensor given by,
\begin{eqnarray}
 T_{AB} = \sigma k_A k_B ,
\end{eqnarray}
where $k_A$ is a radial null 5-vector in the bulk then we may take $u^A$ to be the tangent vector to the brane and normalise so that $T_{AB}u^Au^B = \sigma$, where $\sigma$ is the out-flux of gravitons as measured at the brane.
With this we find that,
\begin{eqnarray}
 T^0_5 &=& \pm \sigma ,\nonumber\\
 T_{55} &=& \sigma .
\end{eqnarray}
\subsection{Brane radiating gravitons}
This model has been considered recently by several authors in both the 
symmetric~\cite{karch,march,lang,Leeper} and asymmetric~\cite{Leeper2,paper6} cases.
For this spacetime we have that $\sigma_\pm = \pm \frac{\alpha}{12}\kappa^2 \rho^2$, where alpha is a 
dimensionless constant, and the evolution of the mass parameters is governed by, 
\begin{eqnarray}
 \dot{m}_\pm =  \left [\frac{\kappa^4}{18}a^4 (\rho + V)  \mp \frac{\Delta m + \frac{\Delta \Lambda}{12} a^4}{\rho + V}  - \frac{\kappa^2}{3} Ha^4 \right ] \frac{\alpha}{12}\kappa^2 \rho^2 ,
\end{eqnarray}
where we have used,
\begin{eqnarray}
\Delta T^0_5 &=& \frac{\alpha}{6}\kappa^2 \rho^2 ,\nonumber\\
\overline{T}_{05} &=& \Delta T_{55} = 0 ,\nonumber\\
\overline{T}_{55} &=& \frac{\alpha}{12} \kappa^2 \rho^2 .
\end{eqnarray}
The equations for the full model become,
\begin{eqnarray}
H^2 &=& \frac{\kappa ^4}{36}\rho (\rho + 2V) -\frac{k}{a^2} + \frac{2\overline{m}(t)}{a^4} + \frac{(12 \Delta m  + \Delta \Lambda a^4)^2}{16 \kappa ^4(\rho + V)^2a^8} ,\nonumber\\
\dot{\rho} &=& - 3H(p + \rho) -\frac{\alpha}{6}\kappa^2 \rho^2 ,\nonumber\\
\dot{\Delta m} &=& \frac{-\frac{\alpha}{6}\kappa^2 \rho^2}{\rho + V}(\Delta m + \frac{\Delta \Lambda}{12}a^4) ,\nonumber\\ 
\dot{\overline{m}} &=&  \frac{\alpha \kappa^4 \rho^2 a^4}{36}(\frac{\kappa^2}{6}(\rho + V) -H) .
\end{eqnarray}
These equations have been studied in \cite{Leeper2}, and in \cite{paper6} where the global nature of the bulk 
spacetimes is also examined.

\subsection{Hawking radiation from bulk black holes}
We now consider the situation of an asymmetric brane model with a black hole in the two bulk spacetimes either side of the brane. We take the black hole mass parameters to be different from one another and include a possible difference in the bulk cosmological constants. The difference in mass parameters provides a contribution to the FRW equation, whose presence would be strong in the early universe. It has previously been assumed that these mass parameters remain constant throughout the model's history, but the black holes can radiate thermally through Hawking radiation and vary the dark radiation and asymmetry terms in time.
Since we assume standard model particles are constrained to the brane, and the bulk is free of any scalar fields, the only channel remaining for Hawking radiation is through bulk gravitons.

We now calculate the evaporation rate for these bulk black holes. We assume that a non-zero 
cosmological constant in the bulk does not substantially affect the evaporation rate.
Following \cite{liddle} we make a spherically symmetric approximation for the black holes and consider the metric,
\begin{eqnarray}
 ds^2 = -f(r)dt^2 + \frac{dr^2}{f(r)} + r^2 d \Omega ^2 ,
\end{eqnarray}
with,
\begin{eqnarray}
 f(r) &=& 1- \frac{r_0^2}{r^2} ,
\end{eqnarray}
For us $r_0 =2m$, which may be related to the ADM mass, $M$ via (see \cite{hemming}),
\begin{eqnarray}
 r_0=2m=\frac{\kappa ^2 M}{3\pi^2} .
\end{eqnarray}
In $D$ dimensions, the rate of decrease of a radiating black hole in the high frequency limit is given by \cite{empar},
\begin{eqnarray}
 \frac{dM}{dt} = - g_D \sigma_D A_D T^D ,
\end{eqnarray}
where $g_D$ is the number of degrees of freedom, $\sigma_D$ is a generalised Stefan's constant, $A_D$ is the geometrical area of the event horizon and $T$ is the temperature of the black hole as measured at infinity.
In 5 dimensions with only graviton emission, this becomes
\begin{eqnarray}
 \frac{dM}{dt} = - \frac{80}{3 \pi^2}\zeta (5)r_0^3T^5 .
\end{eqnarray}
We derive the black hole temperature by considering the periodicity of the Euclidean time for the metric.
The Hawking temperature of the black hole is then given by,
\begin{eqnarray}
 T = \frac{f'(r_0)}{4\pi} = \frac{1}{2 \pi r_0} .
\end{eqnarray}
Consequently we find that the rate of decrease of the mass parameter near infinity is given by,
\begin{eqnarray}
 \left (\frac{dm}{dt} \right )_\infty = - \left ( \frac{80 \zeta (5) \kappa^2}{(6\pi^2)^2(2 \pi)^5 m} \right ) .
\end{eqnarray}
To arrive at the flux of gravitons at the brane we consider, as in section 2.7, a radially outgoing null fluid with the Vaidya metric,
\begin{eqnarray}
 ds^2 = -fdv^2 - 2dvdr + r^2d\Omega^2 .
\end{eqnarray}
The energy momentum in the bulk is again of the form $T_{AB} = \sigma k_A k_B$ which is normalised as,
\begin{eqnarray}
 k_Au^A = 1 ,
\end{eqnarray}
so that $\sigma$ becomes the flux of gravitons measured on the brane. From the Einstein equations we find that,
\begin{eqnarray}
 \frac{dm}{d\tau} = -\frac{\kappa^2r^3 \sigma}{3 \dot{v}} ,
\end{eqnarray}
where $\dot{v} = \frac{dv}{d\tau}$, and
\begin{eqnarray}
 f\dot{v} = \sqrt{\dot{r}^2 + f} - \dot{r} .
\end{eqnarray}
If we now make an identification of these two systems near spatial infinity then we assume that the rate of decrease of the mass function at the brane in the Vaidya spacetime is equal to the derived rate of decrease of an evaporating black hole in AdS-Schwarzschild. Consequently we obtain, after replacing $r$ with $a$ for the scale factor,
\begin{eqnarray}
\frac{\kappa ^2}{3} \sigma = \frac{-(\frac{dm}{dt})(\sqrt{\dot{a}^2 + f} - \dot{a})}{a^3 f } ,\nonumber\\
\sigma = \frac{240 \zeta (5)(\sqrt{\dot{a}^2 + f} - \dot{a})}{a^3 f(6\pi^2)^2(2 \pi)^5 m } ,
\end{eqnarray}
which provides us with the energy flux at the brane from the radiating black hole. We note that this evaporation rate contains a dependence on the brane's motion through the spacetime corresponding to a red-shift of the graviton flow as measured on the brane.

 We may now deal with the model consisting of two black holes, one in each bulk region, and combine together their respective fluxes ( $\sigma_\pm$ ) at the brane. For this situation we can make the further assumption that the outflows and inflows do not interact too strongly with each other and we may analyse the ingoing fluxes and the outgoing fluxes separately. We may then write the total energy flux either side of the brane as,
\begin{eqnarray}
 T^0_{5+} &=&   \sigma_-  - \sigma_+ ,\nonumber\\
 T^0_{5-} &=&   \sigma_-  - \sigma_+ .
\end{eqnarray}
Consequently we have that,
\begin{eqnarray}
\Delta T^0_5 &=& 0 ,\nonumber\\
\overline{T}^0_5 &=&  T^0_{5+,-}=T^0_5 =   \sigma_-  - \sigma_+ ,
\end{eqnarray}
which merely states that the gravitons do not accumulate on the brane.

The pressure terms for this system are given by,
\begin{eqnarray}
 \Delta{T}_{55}= 0 ,\nonumber\\
\overline{T}_{55}=  |\sigma_-  - \sigma_+| .
\end{eqnarray}
We would then obtain the following set of equations,
\begin{eqnarray}
H^2 &=& \frac{\kappa ^4}{36}\rho (\rho + 2V) -\frac{k}{a^2} + \frac{2\overline{m}(t)}{a^4} + \frac{(12 \Delta m  + \Delta \Lambda a^4)^2}{16 \kappa ^4(\rho + V)^2a^8} ,\nonumber\\
\dot{\rho} &=& - 3H(p + \rho) ,\nonumber\\
\dot{\Delta m} &=&  \left (\frac{\kappa^2}{3} \right )^2a^4(\rho + V) T^0_5  ,\nonumber\\
\dot{\overline{m}} &=& \frac{T^0_5}{\rho + V}(\Delta m  + \frac{\Delta \Lambda}{12}a^4) - \frac{\kappa^2}{3}Ha^4 T_{55} .
\end{eqnarray}
We note that the process of radiating black holes tends to \emph{increase} the mass asymmetry across the brane, while the average mass may increase or decrease, depending on the relative sizes of the brane's expansion rate and the bulk asymmetry.

A novel possibility is a model where there is initially a pseudo $Z_2$-symmetry present across the brane 
in that both bulks have the same cosmological constant and mass parameter. However, if only one side 
contained an actual black hole, then the resultant radiation would naturally create an asymmetry by 
its flux of gravitons.

We now pass to dimensionless variables, where we note that the variables possess different dimensions depending on whether $k$ is $0$ or $\pm 1$. 
Using $\hat{X}$ to signify dimensionless equivalent of $X$, our equations 
become,
\begin{eqnarray}\label{dimensionless}
\hat{H}^2 &=& \hat{\rho}(2 + \hat{\rho}) -\frac{k}{\hat{a}^2} + \frac{2\overline{\hat{m}}}{\hat{a}^4} + \frac{(\Delta \hat{m}  + 3 \Delta\hat{\Lambda } \hat{a}^4)^2}{4(\hat{\rho} + 1)^2\hat{a}^8} ,\nonumber\\ 
\frac{d\hat{\rho}}{d\hat{t}} &=& - 4\hat{H}\hat{\rho} ,\nonumber\\
\frac{d\Delta \hat{m}}{d\hat{t}} &=&   24\hat{T}^0_5\hat{a}^4(\hat{\rho} + 1) ,\nonumber\\
\frac{d\overline{\hat{m}}}{d\hat{t}} &=&  \hat{T}^0_5\left [ \frac{6( \Delta \hat{m}  + 36\Delta\hat{\Lambda}\hat{a}^4)}{\hat{\rho} + 1}\right]  - 12\hat{H}\hat{a}^4 \hat{T}_{55} .
\end{eqnarray}
It is also straightforward to include the terms governing graviton production on the brane. The resultant system possesses certain simplifications such as the flow of gravitons being purely radial and the assumption that once energy enters a given bulk it immediately increases the black hole mass. We provide simulation results for this model in the next section.

\section{Numerical analysis}\label{secna}
In this section we present the results from numerically simulating bulk black holes that radiate. We considered the four models obtained by turning off or on the brane flux or the black hole fluxes. It is then possible to determine the separate contributions of each process to the overall cosmological model of a radiating brane moving through a bulk containing radiating black holes. In what follows, we discuss the black hole features separately from the cosmological variables on the brane universe.
\subsection{Black hole flux and mass}
We solved numerically equations (\ref{dimensionless}) with a range of initial black hole masses, $m_+$ and $m_-$. Each simulation was then repeated with a difference in cosmological constants between the two bulk 
spacetimes.

In figure 1 we show the black hole fluxes as measured at the brane for an average black hole mass of 
$\overline{m }=6$ and a difference of $\Delta m =2$. The solid curves ($\sigma_{\pm}$) 
are for the model containing both 
Hawking radiation in the bulk and graviton production on the brane. 
The dotted curves ($\sigma^{bh}{}_{\pm}$) are the results 
without graviton production. We note that the fluxes are only on when the brane is outside the 
event horizon. The resultant motion of the brane causes a reduction in the energy flux of gravitons 
impinging on the brane, as it moves further and further away from the black holes. The fluxes 
$\sigma_+$ and $\sigma_-$ gradually become similar in magnitude and so the variation in the dark 
radiation and asymmetry terms is substantially reduced.
\begin{figure}[!ht]
\begin{center}
\epsfig{file=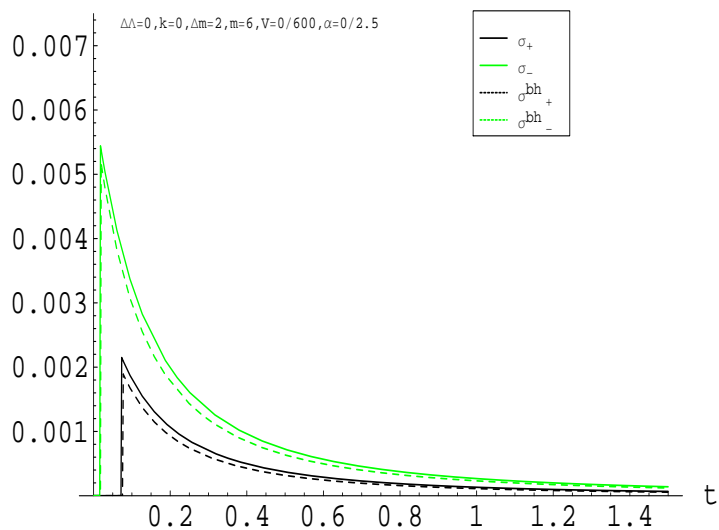 ,height=8cm,width=10cm, angle=0}
\end{center}
\caption{Black hole fluxes according to a brane based observer, initially with
$\Delta m =2$ and $m=6$. The solid lines ($\sigma_{\pm}$) represent the fluxes on the right/left side 
of the brane for the black hole radiating plus brane emitting gravitons case. 
The dotted lines ($\sigma^{bh}{}_{\pm}$) represent the black holes' fluxes 
when the brane does not emit gravitons.}
\end{figure}

For a larger mass difference across the brane, it is easier for the smaller black hole to evaporate completely and the fluxes of the two black holes do not cancel, as is shown in figure 2. The flux of the smaller black hole initially decreases to a minimum and then diverges as it evaporates. We also note that the addition of graviton production on the brane enhances black hole flux in the red-shift phase and diminishes it during the evaporation burst. This effect occurs due to the brane's path, $a(t)$, being perturbed by the gravitons leaking out of the brane. The addition of a small difference in the cosmological constants across the brane results in a slowing down of the previously mentioned behaviours. 
\begin{figure}[!ht]
\begin{center}
\epsfig{file=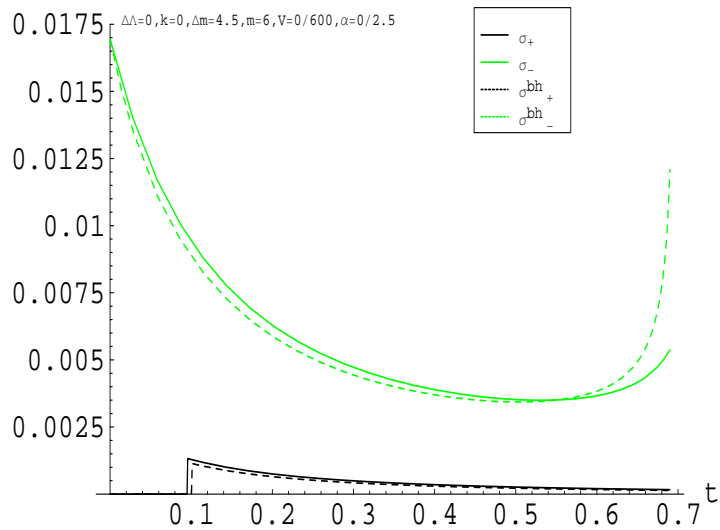 ,height=8cm,width=10cm, angle=0}
\end{center}
\caption{Black hole fluxes according to a brane based observer, initially with 
$\Delta m =4.5$ and $m=6$. Notation is the same as figure 1.}
\end{figure}

The evolution of the average black hole mass, in the brane's frame, is not monotonic in the scenario of pure black hole radiation, as there are competing asymmetry and Hubble terms. Initially, the expansion of the universe dominates over the mass asymmetry and the average black hole mass decreases. From equation (\ref{dimensionfull}) the average mass becomes stationary if,
\begin{eqnarray}
 3 \Delta m + \frac{\Delta \Lambda}{4} a^4&=&(\kappa ^2 \rho_0 + Va^4)H ,
\end{eqnarray}
after which the average mass will increase. The addition of graviton production in the brane does not alter this trend to any great extent, see figure~3. The addition of a small, positive $\Delta \Lambda$ across the brane tends to strongly increase the average black hole mass in the radiating black hole model, but has a minor effect in graviton production model, figure 4.
\begin{figure}[!ht]
\begin{center}
\epsfig{file=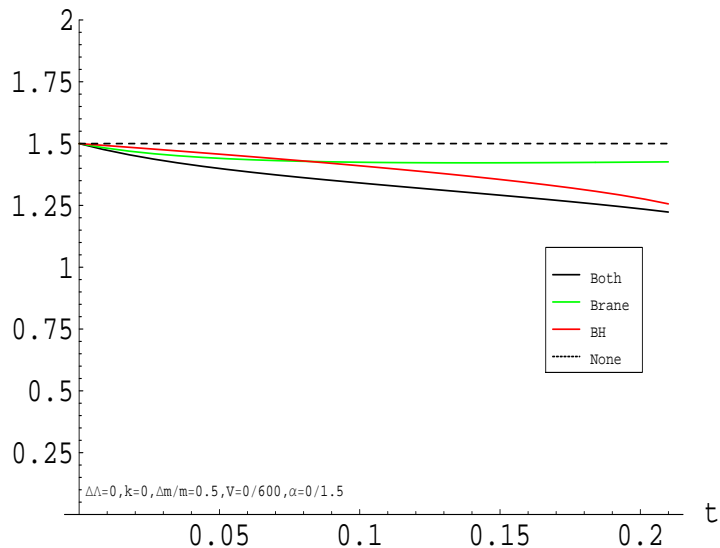,height=8cm,width=10cm, angle=0}
\end{center}
\caption{The behaviour of $\overline{m}(t)$ in the four models. The black solid line represents the 
case where both the brane and the black holes are radiating, the green/red line the case where only the 
brane/black holes are radiating, and the dashed line where no radiation occurs}
\end{figure}
 \begin{figure}[!ht]
\begin{center}
\epsfig{file=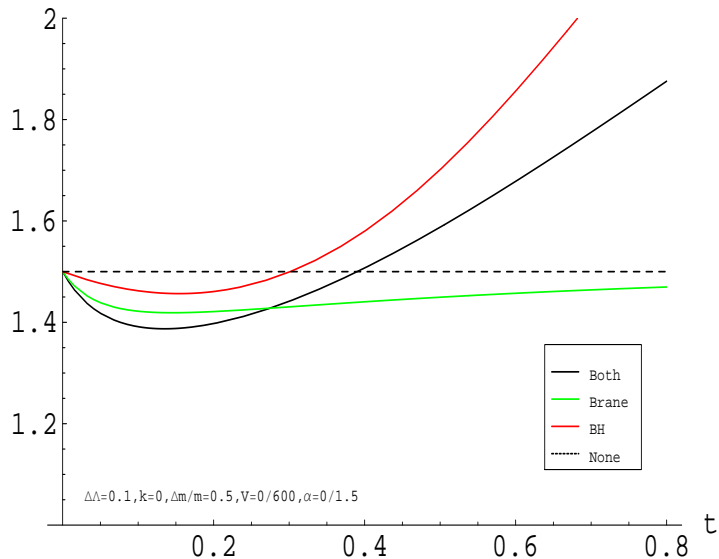,height=8cm,width=10cm, angle=0}
\end{center}
\caption{The behaviour of $\overline{m}(t)$ in the four models with $\Delta \Lambda = 0.1$. Labelling
is the same as figure 3.}
\end{figure}

In the case of pure black hole emission the mass asymmetry always increases in time, as can be seen from equation (\ref{dimensionfull}) this competes against the emission of gravitons from the brane which tends to reduce the asymmetry. Because the flux of gravitons out of the brane depends quadratically on the density of the brane, it can initially dominate over the black hole radiation, but later has little or no affect on the development of the mass asymmetry, figure 5. If we insert a positive difference of bulk cosmological constants in the black hole radiation model the increase in asymmetry is unchanged, as opposed to the radiating brane model where the asymmetry reduces more rapidly, figure 6.
\begin{figure}[!ht]
\begin{center}
\epsfig{file=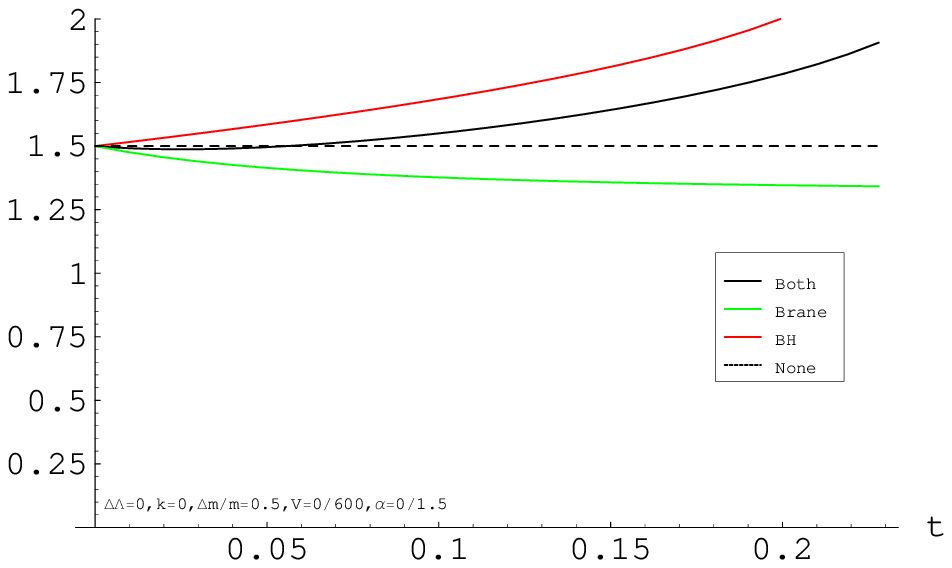,height=8cm,width=10cm, angle=0}
\end{center}
\caption{The behaviour of $\Delta m(t)$ in the four models. Labelling
is the same as figure 3.}
\end{figure}
\begin{figure}[!ht]
\begin{center}
\epsfig{file=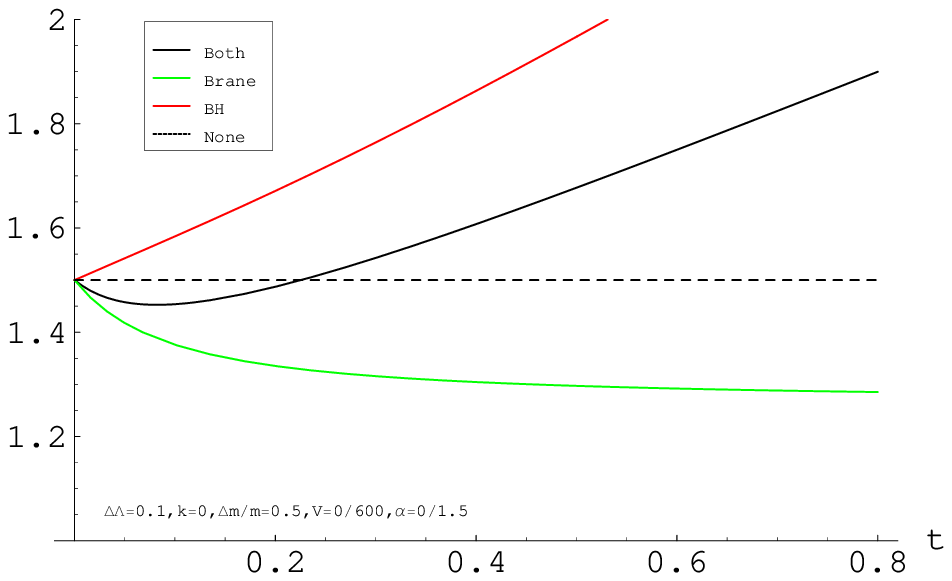,height=8cm,width=10cm, angle=0}
\end{center}
\caption{The behaviour of $\Delta m(t)$ in the four models with $\Delta \Lambda = 0.1$. Labelling
is the same as figure 3.}
\end{figure}

\subsection{Cosmology on the brane}
We now turn to the effect these radiating black holes have on the cosmological parameters in the brane. Their presence manifests itself primarily through the dark radiation and asymmetry terms, and indirectly through the evolution of $\rho (t)$ and $a(t)$.

In figure 7 we have plotted the fractional contribution of the dark radiation term to the FRW equation. We consider its variation over the four models obtained by switching on or off the radiating black holes or the radiating brane. We see that having radiating black holes in the bulk tends to suppress the dark radiation as opposed to the radiating brane, that increases the dark radiation by transfering energy to it from graviton production.  In contrast, radiating black holes tend to magnify the contribution of the asymmetry term as opposed to brane radiation which has little affect, figure 8.
\begin{figure}[!ht]
\begin{center}
\epsfig{file=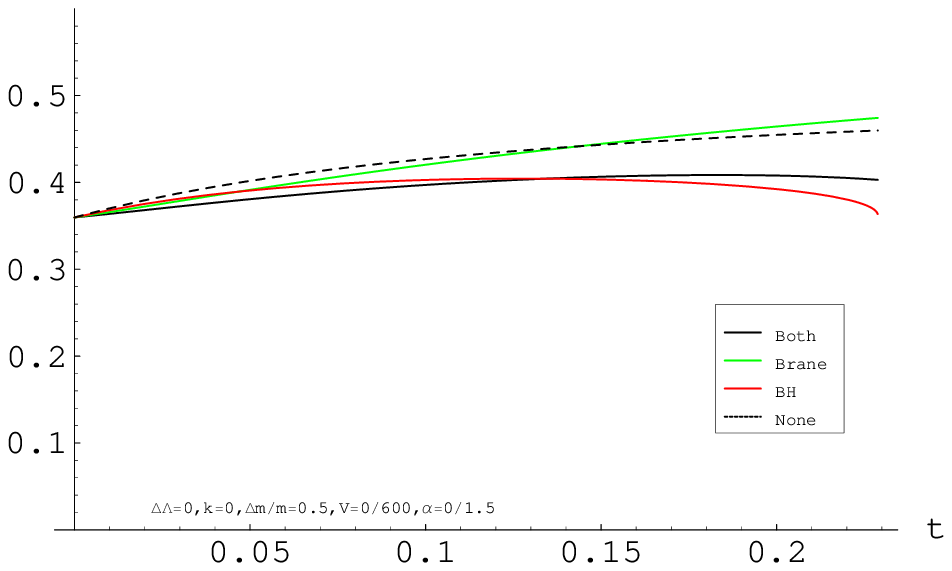,height=8cm,width=10cm, angle=0}
\end{center}
\caption{The fractional contribution of dark radiation to $H^2$. Labelling
is the same as figure 3.}
\end{figure}
\begin{figure}[!ht]
\begin{center}
\epsfig{file=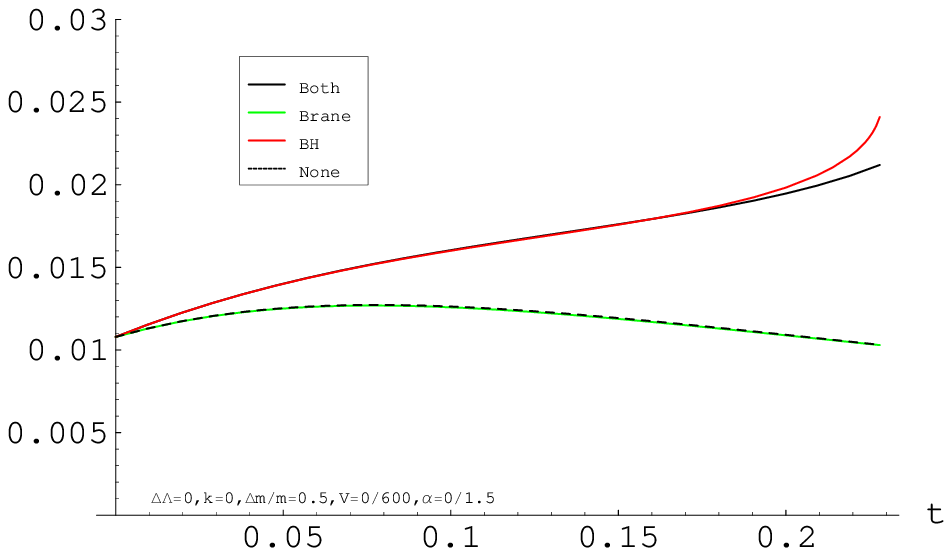,height=8cm,width=10cm, angle=0}
\end{center}
\caption{The fractional contribution of the asymmetry terms to $H^2$. Labelling
is the same as figure 3.}
\end{figure}

The addition of a difference in cosmological constants tends to provide an effective late time cosmological constant in the FRW in the asymmetry term, and is thus constrained experimentally to be small.  It also has the effect of increasing the contribution of dark radiation over time, by enhancing the two graviton processes present.

The scale factor is modified in different ways by the brane and black hole radiation. In general the black hole radiation tends to give a larger scale factor for the universe, while radiation from the brane tends to reduce it. In the presence of a small cosmological constant we may compare the four models in terms of their scale factors, as in figure 9. The Hubble parameter is decreased by both radiation processes, see figure 10, with brane radiation dominating the suppression at early times until the black hole radiation increases in strength.
\begin{figure}[!ht]
\begin{center}
\epsfig{file=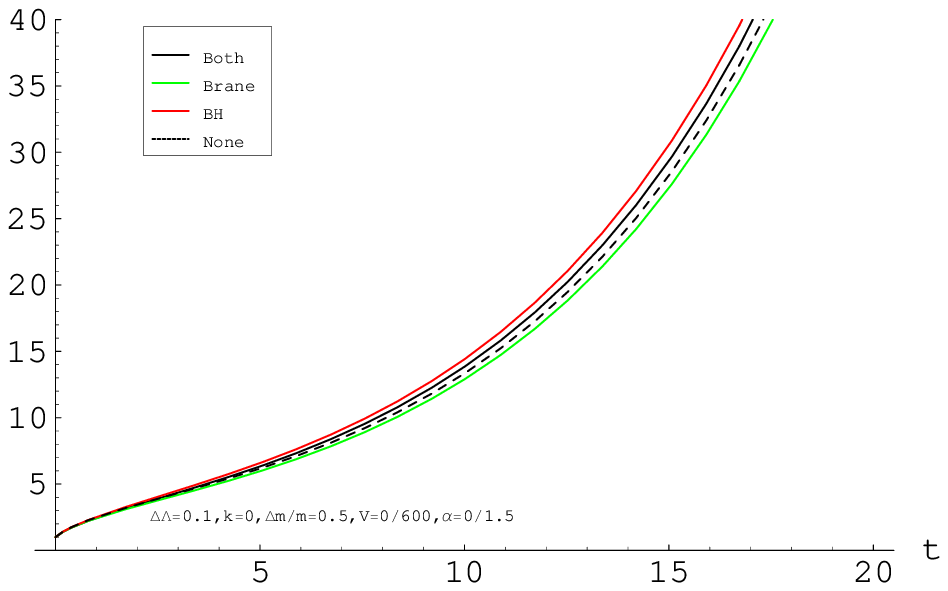,height=8cm,width=10cm, angle=0}
\end{center}
\caption{The variation of the scale factor in the four models. Labelling
is the same as figure 3.}
\end{figure}
 \begin{figure}[!ht]
\begin{center}
\epsfig{file=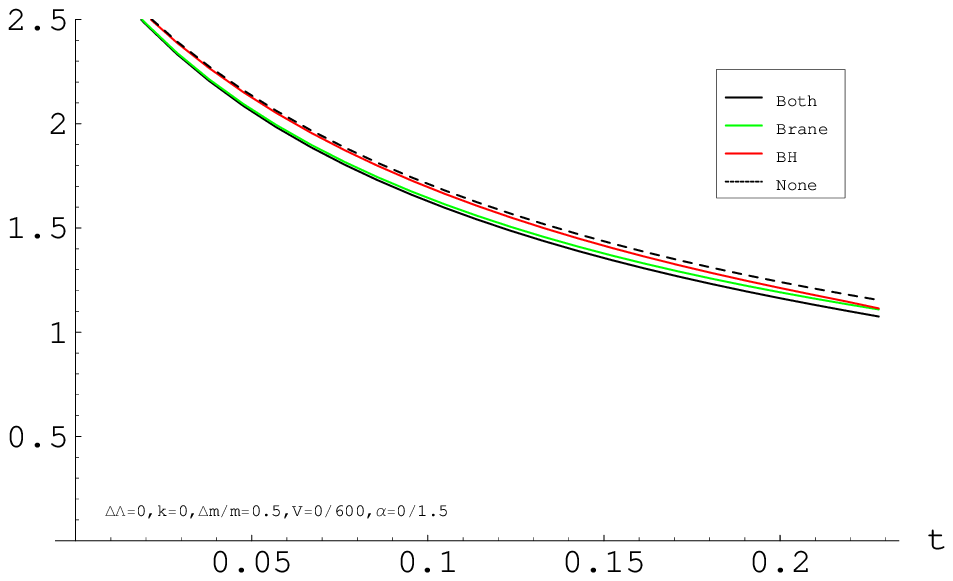,height=8cm,width=10cm, angle=0}
\end{center}
\caption{The variation of the Hubble parameter in the four models. Labelling
is the same as figure 3.}
\end{figure}

\section{Conclusion}\label{secc}
In this paper we developed the general framework necessary to describe an asymmetric brane universe with arbitrary bulk content. We highlighted the fact that in the absence of $Z_2$-symmetry, it is necessary to include a passive source equation, describing the motion of the brane within the background spacetime. Through the FRW equation we were able to introduce a generalised dark radiation term, which we showed was composed of two natural pieces. One piece is the familiar dark radiation term, while the other contained the asymmetry terms for the model. Their evolution in time indicate that the expressions $\frac{3}{2 \kappa ^2}C \pm \frac{1}{2}F$ are intrinsic to the two different bulks and a calculation in a Vaidya spacetime revealed that they correspond to the mass parameters. We then arrived at a locally defined mass parameter which generalised this identification with the dark radiation and asymmetry terms in the FRW equation. 

In section 3 we retricted our equations to two cases of particular interest: graviton production on the brane and radiating bulk black holes. We calculated the flux of gravitons from these bulk black holes as measured on the brane and performed numerical simulations of the system. It was found that radiating black holes tend to increase the asymmetry term in time, while the brane radiation reduces it. The effect on the dark radiation was less straightforward with a competition between cosmological expansion and asymmetry terms.

The two radiation processes produce corrections to the scale factor and the Hubble parameter (figures 9 and 10), while the black hole radiation amplifies the asymmetry term in the FRW equation. 
The analysis we have presented can be used to determine the values of $\overline{m}$ and $\Delta m$ at the 
time of nucleosynthesis, and thereby to impose strict constraints on the initial conditions of the 
system, but we leave this to future work.
Other interesting investigations in this area could include an analysis of non-radial graviton flows 
and the effects associated with having scalar fields present in the bulk.

\bibliographystyle{JHEP} \bibliography{thesis1}






\end{document}